\documentclass[prb,twocolumn]{revtex4}

\usepackage{amsmath}
\usepackage{graphicx}
\begin{document}

\title{What Brown saw and you can too}

\author{Philip Pearle}
\email{ppearle@hamilton.edu}
\affiliation{Department of Physics, Hamilton College, Clinton, New York 13323}

\author{Kenneth Bart}
\affiliation{Department of Biology, Hamilton College, Clinton, New York 13323}

\author{David Bilderback}
\affiliation{Division of Biological Sciences, The University of Montana, Montana 59812}

\author{Brian Collett}
\affiliation{Department of Physics, Hamilton College, Clinton, New York 13323}

\author{ Dara Newman}
\affiliation{Division of Biological Sciences, The University of Montana, Montana 59812}

\author{Scott Samuels}
\affiliation{Division of Biological Sciences, The University of Montana, Montana 59812}

\begin{abstract}
A discussion is given of Robert Brown's original observations of particles ejected by pollen of the plant \textit{Clarkia pulchella} undergoing what is now called Brownian motion. We consider the nature of those particles, and how he misinterpreted the Airy disc of the smallest particles to be universal organic building blocks. Relevant qualitative and quantitative investigations with a modern microscope and with a ``homemade" single lens microscope similar to Brown's, are presented.
\end{abstract}

\maketitle

\section{Introduction}\label{S1}

In June 1827 the British botanist Robert Brown used his one lens microscope (a magnifying glass with a small diameter and large curvature) to observe pollen of the plant \textit{Clarkia pulchella} immersed in water. He noticed that particles ejected from the pollen were of two shapes: some were oblong and some of the smaller ones were circular and moved at random in the water. Thus commenced his investigations, which showed that anything sufficiently small would move similarly. We now understand, as Brown did not, that the motion is due to the random impact of water molecules.

This paper arose in part from curiosity as to the nature of the particles that Brown observed. The oblong particles Brown saw are amyloplasts (starch organelles, that is, starch containers) and the spherical particles are spherosomes (lipid organelles, that is, fat containers). The nature of the particles (although not their current names) was known in 1848.\cite{LindleyBook}

Brown was motivated by his observation that the smallest bits in motion were circular and about the same diameter. He called these bits ``molecules" (a word then in common usage meaning tiny particle), suggesting that they might be universal building blocks of organic matter. What Brown actually saw was the diffraction ring (Airy disc) and possible spherical aberration caused by his lens when he viewed sufficiently small objects. We know of only one article\cite{vanderpas0} which questions the effect of Brown's lens.

The aim of this paper is to provide a quantitative understanding using modern equipment of Brown's observations and to enable the reader to perform similar experiments.

Information is provided on pollen physiology and on how to build a single ``ball" lens microscope. Amyloplast and spherosome sizes, seen with an electron microscope, a modern compound microscope, and the ball lens microscope are compared with each other and with Brown's measurements, providing an understanding of the effect of the Airy disc on the optical observations. Measurements of Brownian motion of amyloplasts and spherosomes are compared with theory.

An extended treatment of sections of this paper, along with instructions on how to grow \textit{Clarkia pulchella} and prepare its pollen for observation, and mathematical tutorials on the viscous force felt by particles and on single lens optics (including a unified treatment of diffraction and spherical aberration), and additional materials such as movies of pollen bursting and emitting particles, is available.\cite{website} 

\section{History}

The plant \textit{Clarkia pulchella} was discovered on the return trip of the Lewis and Clark expedition by Meriwether Lewis on June 1, 1806. Lewis wrote: ``I met with a singular plant today in blume, of which I preserved a specemine. It grows on the steep sides of the fertile hills near this place.\ldots\ I regret very much that the seed of this plant are not yet ripe and it is probable will not be so during my residence in this neighborhood."\cite{MLewis}

Upon returning, Lewis hired Frederick Traugott Pursh in 1807 to prepare a catalog of the plants he had collected. Lacking support from Lewis, Pursh sailed to London in the winter of 1811, taking his work and many of Lewis's specimens with him. Pursh published his volume \textit{Flora Americae Septentrionalis} in mid-December 1813. In it, he gave the name \textit{Clarkia pulchella} (beautiful Clarkia) to the flower in honor of Clark.

Seeds of \textit{Clarkia pulchella} were available in England only in 1826, through an expedition to the American northwest begun on July 25, 1824, sponsored by the Horticultural Society of London, and by the Hudson's Bay Company, undertaken by David Douglas (after whom the Douglas fir is named). Douglas wrote in his journal\cite{Douglas} a list and description of the plants he found in July, 1825, including: ``\textit{Clarkia pulchella} \textit{(Pursh)}, annual; description and figure very good; flowers rose color; abundant on the dry sandy plains near the Great Falls; on the banks of two rivers twenty miles above the rapids; an exceedingly beautiful plant. I hope it may grow in England." He shipped the seeds on October 25, 1825, and they arrived at London on April 15, 1826.\cite{Quarry} The pollen grown the following year from \textit{Clarkia pulchella} plants was put to use by Brown.

Robert Brown\cite{Mabberley} was a proteg\'{e} of Joseph Banks, the most eminent botanist of his time. In 1800 Banks offered Brown the post of naturalist on an Admiralty sponsored expedition to the coast of Australia. Brown embarked on the ship \textit{Investigator} on July 18, 1801 and returned on October 7, 1805. He found thousands of new species of plants and eventually achieved renown comparable to Banks.

Banks died in mid-1820 and bequeathed his library, herbarium, an annuity and the lease to his house to Brown, with the stipulation that Brown take up residence there. While negotiating with the Trustees of the British Museum for the transfer of Banks's library, concluding in September 1827, Brown conducted the investigations we will discuss.

\section{Brown's Investigations}

In Brown's ``Slips Catalogue," dated June 12, 1827 and June 13, 1827, are some sheets labeled \textit{Clarkia}.\cite{slips} Directly underneath, Brown's (not always legible) handwriting reads \textit{Hort Soc} (Horticultural Soc) \textit{Horticult (illegible) Chiswick (illegible)}. The next line reads \textit{occident }(western) \textit{Amer (illegible) by D Douglas}. That is, Brown certifies that his \textit{Clarkia pulchella} flowers came directly from the Horticultural Society's garden in Chiswick.

\begin{figure}[h]
\begin{center}
\includegraphics[width=0.5\textwidth]{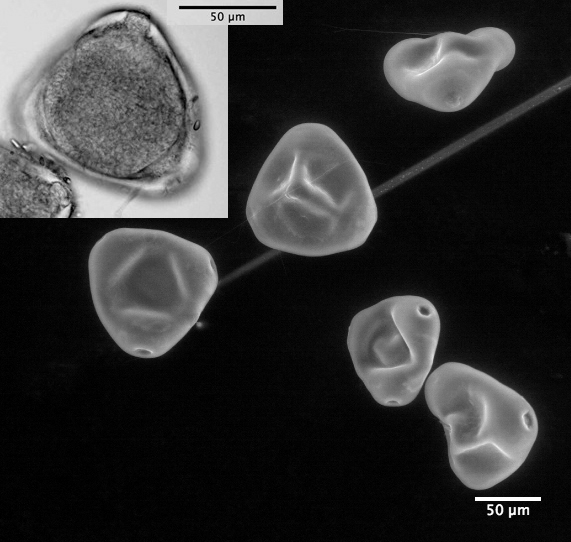}
\caption{\label{f2} \textit{Clarkia pulchella} pollen imaged by a microscope (inset) and by an electron microscope.}
\label{default}
\end{center}
\end{figure}

The first entry describes the pollen: ``The grains of Pollen are subspherical or orbiculate-lenticular with three equidistant more pellucid and slightly projecting points so that they are obtusely triangular \ldots\ ."
Figure~\ref{f2} is an electron microscope picture and the inset is an optical microscope picture of the pollen. They look vaguely like pinched tetrahedrons, with the longest dimension around $100\,\mu$m and ``pores" at three vertices.\cite{Small}

The entry then describes the pollen contents: ``The fovilla or granules fill the whole orbicular disk but do not extend to the projecting angles. They are not spherical but oblong or nearly cylindrical. \& the particles have a manifest motion. This motion is only visible to my lens which magnifies 370 times. The motion is obscure but yet certain \ldots."

Thus began the research that resulted in Brown's paper,\cite{Brown} dated July 30, 1828. It was first circulated as a pamphlet, and then published in September 1828. The first paragraph describes his microscope: ``The observations, of which it is my object to give a summary in the following pages, have all been made with a simple microscope, and indeed with one and the same lens, the focal length of which is about 1/32 of an inch." A well known rule of thumb is that a near object is best seen at a distance of 10 inches. This rule puts the magnification Brown used at $\approx 10/f=320$, which is not far from Brown's own estimate of $\times 370$.

It has been conjectured that the two extant microscopes of Brown's, one at Kew Gardens, the other at the Linnean Society, represented his full collection, and that the latter's $\times 170$ lens was the one used for the Brownian motion observations.\cite{fordb,fordL} Both these conjectures are doubtful.\cite{refuteford}

The purpose of Brown's investigation was to ``attend more minutely than I had before done to the structure of the Pollen \ldots" and to determine its mechanism of fertilization. He first looked at \textit{Clarkia pulchella} pollen because he had some and could see oblong particles within, which he thought would best allow him to follow their motion during the fertilization process.\cite{note}

Brown next wrote that ``This plant was \textit{Clarckia pulchella}, of which the grains of pollen, taken from antherae fully grown before bursting, were filled with particles or granules of unusually large size, varying from 1/4000th to about 1/3000th of an inch in length, and of a figure between cylindrical and oblong \ldots\ . While examining the form of these particles immersed in water, I observed many of them very evidently in motion \ldots\ ." This particle was the first observed by Brown, whose length he estimated to be $\approx 6\ \mbox{to}\ 8 \times 10^{-3}$\,mm. (We will find that these particles have shorter lengths. The difference shall be attributed to the alteration of the image by his lens.)

Brown noted a second kind of particle: ``Grains of pollen of the same plant taken from antherae immediately after bursting, contained similar subcylindrical particles, in reduced numbers however, and mixed with other particles, at least as numerous, of much smaller size, apparently spherical, and in rapid oscillatory motion.

These smaller particles, or Molecules as I shall term them, \ldots\ on continuing to observe them until the water had entirely evaporated, both the cylindrical particles and spherical molecules were found on the stage of my microscope."

\subsection{Brownian Motion}
We emphasize that Brown did not observe the pollen move. Instead, he observed the motion of much smaller objects that reside within the pollen.\cite{Layton} Nonetheless, statements that Brown saw the pollen move are common.\cite{Bernstein}

A \textit{Clarkia} pollen is $\approx 100\,\mu$m across,\cite{Small} which is too large for its Brownian motion to be readily seen. Fortunately for Brown, the contents of the pollen are the right size for their motion to be conveniently observed. To obtain a rough understanding of what may not or may be seen undergoing Brownian motion, we employ Eq.~(\ref{A3}) for the mean square distance traveled by a particle in time $t$ in one dimension in a liquid at temperature $T$ which provides a velocity-dependent viscous force $-\beta v$:
\begin{equation}
\overline{x^{2}}=\frac{2kTt}{\beta},
\end{equation}
where $k$ is Boltzmann's constant. As shown in Eq.~(\ref{A8}), the mean distance traveled is $\overline{|x|}\approx 0.80\sqrt{\overline{x^{2}}}$.

For a sphere of radius $R$, Stokes law gives $\beta=6\pi\eta R$, where $\eta$ is the (dynamic) viscosity of the liquid. For an oblong object, $R$ is replaced by an effective radius $R_{\rm eff}$, which depende on the angle between the direction of motion and the long axis.

Similar results hold even for a weirdly-shaped object like \textit{Clarkia} pollen. For the pollen and its contents we write
\begin{equation}\label{1}
\overline{|x|} \approx 0.80\sqrt{\frac{2kTt}{6\pi\eta R_{\rm eff}}}. 
\end{equation}

\begin{table}[h!]
\centering
\begin{tabular}{|l||l|l|l|c|c|c|r|r|r|r|}
\hline
$R_{\rm eff}\,(\mu{\rm m})$ & 0.50&1.0&1.5&2.0&2.5&3.0&3.5&4.0&50\\ \hline\hline
$t=1$& 0.74& 0.52& 0.43& 0.37& 0.33& 0.30& 0.28 & 0.26& 0.07\\ \hline
$t=30$ &4.1&2.9&2.3&2.0&1.8&1.6&1.5&1.4&0.41\\ \hline
$t=60$ &5.7&4.0&3.3&2.9&2.6&2.3&2.2&2.0&0.57\\
\hline
\end{tabular}
\caption{Average mean distance $\overline{|x|}$ ($\mu m$) traveled under Brownian motion, in water at temperature $T=20^{\circ}$C, for various times $t$ in seconds, for an object whose size is characterized by $R_{\rm eff}$. } 
\end{table}

Table I follows from Eq.~(\ref{1}). The reason for choosing $t=1$\,s is that the little jiggles on the time scale of about a second are what catches the eye. For later use, Table~II provides the results for the mean angle $\overline{|\theta|}$ given by Eq.~(\ref{A9}):
\begin{equation}\label{2}
\overline{|\theta|}\approx 0.80\sqrt{\frac{2kTt}{8\pi\eta R_{\rm eff}^{3}}},
\end{equation}
where $R_{\rm eff}$ depends on the axis of rotation. 

\begin{table}[h!]
\centering
\begin{tabular}{|l||l|l|l|c|c|c|r|r|r|r|}
\hline
$R_{\rm eff}\,(\mu{\rm m})$ &0.50&1.0&1.5&2.0&2.5&3.0&3.5&4.0&50\\ \hline\hline
$t=1$ &74&26&14&9&7&5&4&3&0.01\\ \hline
$t=30$ &402&142&78&50&36&27&22&18&0.4\\ \hline
$t=60$ &570&201&110&71&51&39&31&25&0.6\\
\hline
\end{tabular}
\caption{Average mean angle $\overline{|\theta|}$ in degrees 
 traveled under Brownian motion, in water at temperature $T=20^{\circ}$C, for various times $t$ in seconds, for an object whose size is characterized by $R_{\rm eff}$. } 
\end{table}

A rough argument as to why the pollen contents, but not the pollen, can be seen to undergo Brownian motion goes as follows. The human eye is considered unable to resolve angles less than 1 arcminute $\approx 2.9\times 10^{-4}$\,rad.\cite{eye} At a distance of 25\,cm, this limit means that a displacement less than $73\,\mu$m cannot be seen by the eye. Thus, a displacement less than $73/370\approx 0.2\,\mu$m cannot be seen by the eye with the help of a lens of magnification $\times$370. Therefore, according to this crude hypothesis\cite{crude} based on the Rayleigh criterion, if the eye cannot resolve two separate points, it cannot see motion of an object whose center travels between these points. It follows from Table~I that the pollen with $R_{\rm eff}<4\,\mu$m could be seen to move in 1\,s, but not the pollen with $R_{\rm eff}\approx 50\,\mu$m.

\subsection{Further Observations}\label{S3D}

Brown next looked at the pollen of other plants and saw that their contents and behavior are similar. Up to this point, Brown had not observed the particles or granules moving while they were within the \textit{Clarkia pulchella} pollen grain. As he wrote, he observed them moving in water. Unfortunately, he did not say how the particles get out of the pollen grain after the grains are put in water.

As will be discussed in Sec.~\ref{S4}, pollen grains in water (in vitro) may burst open, the contents streaming out under pressure (called turgor). Moreover, the particles within \textit{Clarkia pulchella} pollen seem to be too packed together to move. We have observed that the fluid in which they are packed is so viscous that their motion is impeded when they do emerge.

However, Brown then says that he was able to see particles move within the pollen of some plants other than \textit{Clarkia pulchella} or its family. Sometimes Brown is said to have observed particles moving within the pollen, and the implication is that this was what he first observed, which is incorrect.\cite{ford}

Then, Brown describes an accident. On ``bruising a spore of \textit{Equisetum}, \ldots which at first happened accidentally, I so greatly increased the number of moving particles that the source of the added quantity could not be doubted."
This increase led him to ``bruise \ldots all other parts of those plants \ldots," with the same motion observed. Therefore, the motion had nothing to do with plant reproduction: ``\ldots\ My supposed test of the male organ was therefore necessarily abandoned."

The naturalist George-Louis Leclerc, Comte de Buffon, had proposed the hypothesis that there are elementary ``organic molecules" (hence Brown's name for the smaller particles he observed) out of which all life is constructed. Brown first subscribes to this hypothesis, because he found by ``bruising" any part of the plants, ``\ldots I never failed to disengage the molecules in sufficient numbers to ascertain their apparent identity in size, form, and motion, with the smaller particles of the grains of pollen."

But, then he commenced to ``bruise" non-organic matter, glass, rocks, minerals, even ``a fragment of the Sphinx," and ``\ldots in a word, in every mineral I could reduce to a powder sufficiently fine to be temporarily suspended in water, I found these molecules more or less copiously \ldots\ ." Therefore, the hypothesis was abandoned.

\subsection{Brown's Summary of Observations on Molecules} \label{S3G}

Brown summarized, with commendable caution:
``There are three points of great importance which I was anxious to ascertain respecting these molecules, namely, their form, whether they are of uniform size, and their absolute magnitude. I am not, however, entirely satisfied with what I have been able to determine on any of these points.
As to form, I have stated the molecule to be spherical, and this I have done with some confidence." He explained that he judged the size of bodies ``\ldots\ by placing them on a micrometer" (a glass slide with lines ruled on it) ``divided to five thousandths of an inch \ldots\ The results so obtained can only be regarded as approximations \ldots\ I am upon the whole disposed to believe the simple molecule to be of uniform size \ldots\ its diameter appeared to vary from 1/15,000 to 1/20,000 of an inch." Therefore, using his microscope, he estimated the molecule size to be from 1.7 to 1.3$\,\mu$m. Brown prudently concluded, ``I shall not at present enter into additional details, nor shall I hazard any conjectures whatever respecting these molecules \ldots\ ." He mentioned people to whom he showed the motion phenomenon and people who had made earlier related observations (it was first seen by Leeuwenhoek, and remarked upon by some later microscopists\cite{vanderpas0}).

Brown issued an addendum the following year.\cite{Brown} He rejected the notion that the molecules are animated, he regretted having introduced hypotheses such as larger objects being made out of molecules, distanced himself from the notion that the molecules are identically sized, and rejected some explanations of the motion. He says they are ``\ldots motions for which I am unable to account."

\section{Botany}\label{S4}

Because this paper concerns observations of the contents of \textit{Clarkia pulchella} pollen, we summarize what was known then and now about pollen and its contents.

Unknown to Brown when he undertook this work in 1827, the mechanism of fertilization of the ovule by pollen had been observed by accident in 1822 by the optical designer, astronomer and botanist Giovanni Battista Amici. Amici was looking at a grain of pollen that had fallen onto a stigma,\cite{Amici} and observed that a tube emerged from the pollen and traveled down the style: ``\ldots I was quite surprised to see it filled with small bodies, part of which came out of the grain of pollen \ldots\ ."

Brown eventually became aware of Amici's discovery of the pollen tube, and was the first to realize that contact with the stigma causes the pollen to germinate. If,  in 1827,  Brown had decided to observe pollen in vivo instead of in vitro, he likely would have seen the pollen tube, pursued that, and his Brownian motion observations might never have taken place.

A summary of our present understanding follows. A pollen grain consists of a cell wall surrounding a single living cell,\cite{1} called a tube cell, because it can grow into a pollen tube. The wall is often optically opaque. (The oblong amyloplasts of \textit{C. pulchella} can be seen through the pollen wall, which led Brown to start working with this plant.) At certain locations there are one or more apertures or pores in the cell wall (\textit{C.\ pulchella} has three pores). When pollen lands on a flower's stigmatic surface, the pollen absorbs water through the pores and various molecules\cite{3,4} of the stigma induce the pollen to germinate, with a pollen tube emerging through one of these pores. When pollen grains are more uniformly surrounded by an artificial incubation medium, several tubes may emerge from a single pollen grain.

Within the tube cell,\cite{5} there is a centrally located nucleus (the cell nucleus was first discovered by Brown) surrounded by the cytoplasm, which consists of a viscous fluid and its contents, membrane-bound structures called organelles. These organelles include amyloplasts (which store starch) and spherosomes (which store lipids) and numerous very small ribosomes needed for protein synthesis. The nature of amyloplasts and spherosomes became known within a decade after Brown's publications, but at that time much was still unknown and wrongly conjectured about plant fertilization\cite{Lindleyquote}.

There is sequential synthesis of cellular components, beginning with starch accumulation in the amyloplasts, then ribosome generation, followed by lipid synthesis in the spherosomes: this sequence underlies Brown's observation that the spherosomes were absent before dehiscence,\cite{note} but visible in greater numbers after dehiscence.

All these cellular components, as well as a generative cell (responsible for fertilization), are passively transported within the elongating pollen tube. The generative cell ultimately divides to form two non-motile sperm, and when the pollen tube, passing through the stigma and down the style, reaches an ovule, the two sperm are released.\cite{6} The egg is fertilized to form the embryo (seedling), and the other sperm unites with the ``central cell" it finds there to form the endosperm, which will become the ``food'' for the seedling. The starch and lipid, stored in the amyloplasts and spherosomes, respectively, are presumably utilized as energy sources and provide raw materials for the construction of new pollen tube wall material during pollen tube elongation.

When pollen grains of many plants are placed in water, they frequently rupture to release the cytoplasmic contents of the tube cell into the water. As the cytoplasmic contents disperse into the water, the more numerous and larger amyloplasts and spherosomes are seen. Other organelles are too small (ribosomes are $\approx 0.02\,\mu$m) to be seen with a light microscope or too few (the nucleus and generative cell) to be easily spotted.
For this reason, when Brown put pollen into water, he saw just the amyloplasts and spherosomes, and thereby discovered Brownian motion.

\section{microscopy}\label{S5}

\textit{Clarkia pulchella}, variously called ragged robin, pinkfairies, elkhorn and deerhorn (because of its four three-pronged petals), is native to western North America.\cite{Lewis}
Several companies sell \textit{Clarkia pulchella} seeds.\cite{seeds} We have put advice on-line\cite{website} for growing \textit{Clarkia} flowers, harvesting the pollen and preparing a slide for viewing.  

\begin{figure}[b]
\begin{center}
\includegraphics[width=0.5\textwidth]{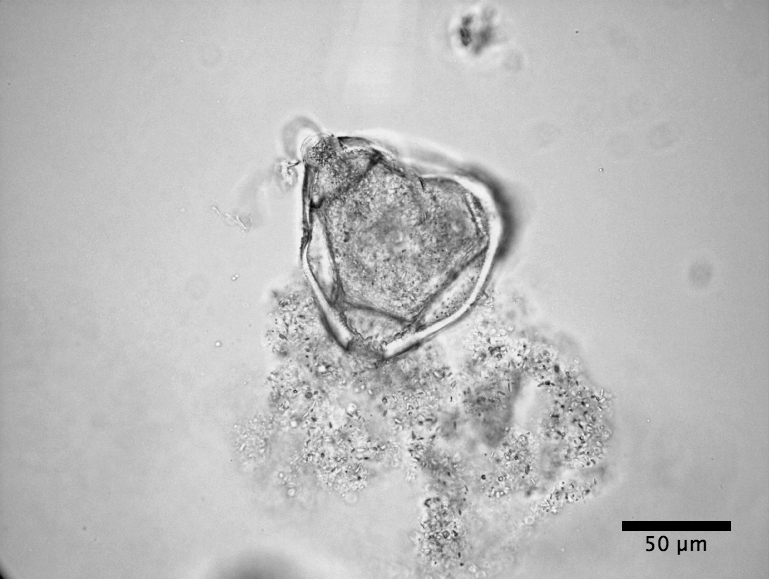}
\caption{\label{f7}Bursting \textit{Clarkia pulchella} pollen.}
\label{default}
\end{center}
\end{figure}
	
\begin{figure}[h]
\begin{center}
\includegraphics[width=0.5\textwidth]{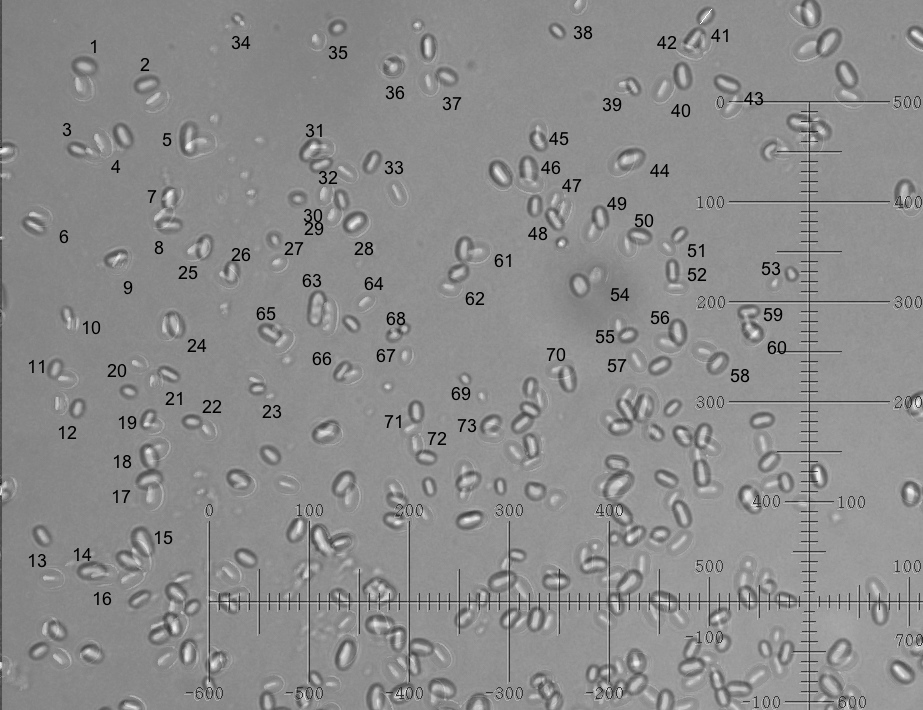}
\caption{\label{f9}\textit{Clarkia pulchella} pollen contents before dehiscence, two superimposed photos taken 1 min apart. The scale is 2 $\mu$m per division.}
\label{default}
\end{center}
\end{figure}

Pollen can be seen bursting, with amyloplasts and spherosomes streaming (Fig.~\ref{f7}) like logs released from a log jam. Particles at the log jam periphery diffuse away from the rest and can be seen undergoing Brownian motion. The remainder are packed closely, and the intracellular medium in which they sit is viscous, and thus they show little or no Brownian motion until the log jam disperses.

We shall discuss three brief studies of particle sizes and their motion. We first consider the distribution of particle sizes emerging from \textit{Clarkia pulchella} pollen before and after dehiscence, verifying Brown's observation that  spherosomes are more numerous after dehiscence. Then we explain the uniform spherosome sizes observed by Brown as due to the Airy disc caused by his lens. Although Brown did no quantitative study of Brownian motion and Brownian rotation of amyloplasts, we shall do so and make a rough comparison of observation with theory. Finally, we exhibit again the effect of diffraction on image size by showing that amyloplast sizes observed with an electron microscope and with a commercial optical microscope are smaller than the sizes reported by Brown and observed with a single lens microscope of power comparable to Brown's, whose construction is discussed in Appendix \ref{SD}.

\subsection{Data on Particle Sizes and their Motion}

An Olympus BX-50 microscope at $\times 400$ was used for our observations. Its resolution is 0.45\,$\mu$m, and its depth of focus is 2.5\,$\mu$m. A microscope camera and five different computer applications were employed,.

Figure~\ref{f9} shows two superimposed photos of \textit{Clarkia pulchella} particles taken 1\,min apart from pollen before dehiscence. The two pictures were enhanced in contrast and treated differently in brightness and then superimposed, using Photoshop Elements 2. A free program, ImageJ,\cite{ImageJ} was used to make precision measurements; 73 particles in the upper left quadrant of the viewing area (two time-displaced images of each) were labeled. Each image's long axis length, long axis angle $\theta$, and $x$ and $y$ coordinates were measured.

\begin{figure}[h]
\begin{center}
\includegraphics[width=0.5\textwidth]{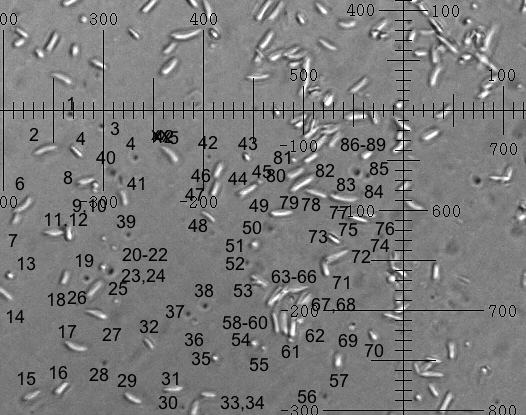}
\caption{\label{f30}\textit{Clarkia pulchella} pollen contents after dehiscence. The scale is 2 $\mu$m per division.}
\label{default}
\end{center}
\end{figure}

\begin{figure}[h]
\begin{center}
\includegraphics[width=0.5\textwidth]{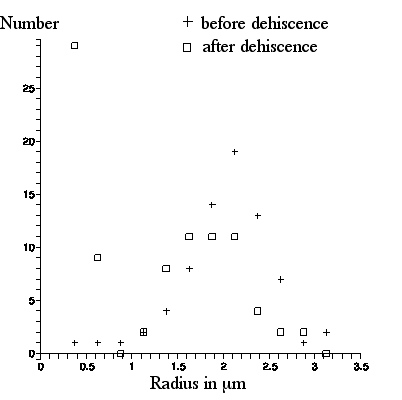}
\caption{\label{f31}Number of particles  (in a radius bin .25 $\mu$m wide) vs radius in $\mu$m.}
\label{default}
\end{center}
\end{figure}

Figure~\ref{f30} shows a photo of \textit{C. pulchella} particles from pollen after dehiscence. 89 particles in the lower left quadrant were labeled and their lengths were measured. A plot of the distribution of their lengths for both photos is in Fig.~\ref{f31}. Note that the radius $R$ is defined as 1/2 the length of the long axis of the amyloplasts. The photos and the graph qualitatively confirm what Brown wrote.

There are few spherosomes visible in Fig.~\ref{f9}, taken from pollen before dehiscence, but Fig.~\ref{f30} shows many more after dehiscence. These particles appear as light or dark, depending on their location with respect to the microscope focal plane.

From Fig.~\ref{f31} we can see that the distribution of numbers of particles with radii larger than $1\,\mu$m before and after dehiscence appears to be the same: these are the amyloplasts. After dehiscence, there is a sharp peak in the number of particles with radii less than $1\,\mu$m (and no such peak before dehiscence): these are the spherosomes.

Quantitatively, there is a discrepancy between Brown's observation of the sizes of the amyloplasts and spherosomes and what is depicted in Fig.~\ref{f31}: his sizes are larger. As we have noted, Brown claimed the amyloplasts to have an average radius (half the long axis length) of $R\approx 3\,\mu$m, with maximum $R\approx 4\,\mu$m. From Fig.~\ref{f31} our Olympus microscope gives, on average, $R\approx 2\,\mu$m, with maximum $R\approx 3\,\mu$m. Brown wrote that the spherosome radii ranged from $R\approx 0.65\,\mu$m to $R\approx 0.85\,\mu$m. From Fig.~\ref{f31} most spherosomes appear to cluster around $R\approx 0.5\pm 0.05\,\mu$m, with a maximum of $R\approx 0.65\pm 0.05\,\mu$m.

\subsection{Effect of Lens on Observed Size}

Brown was stimulated in writing his paper by his observations and hypothesis of the ubiquity and uniformity of the ``molecules." As mentioned in Sec.~I, when he viewed objects smaller than the resolution of his lens, diffraction and possibly spherical aberration produced a larger, uniform size.\cite{vanderpas} We now discuss this point further, summarizing results given in Appendix~\ref{SC}. The main result is Fig.~\ref{F32}, which enables us to find the actual radius $a$ of a spherical object from the larger radius $R$ of the image observed through a lens or microscope. From information supplied by Brown about the size of his ``molecules," we can guess the radius of the circular aperture (the exit pupil) which backed his microscope lens.

For sufficiently small radius $b$ of the exit pupil of a lens, the image of a point source of light is a circular diffraction pattern. The intensity as a function of radial distance $r$ from the lens axis is given by
\begin{equation}\label{3}
I_{A}(x)=\Bigg[\frac{2J_{1}(x)}{x}\Bigg]^{2},
\end{equation}
where $J_{1}(x)$ is the Bessel function (the normalization is taken so that $I_{A}(0)=1$) and $x\equiv krb/f$, where $f$ is the lens focal length, $k=2\pi/\lambda$, and $\lambda$ is the wavelength of the light, traditionally taken for design purposes as green with $\lambda= 0.55\,\mu$m. Equation~\eqref{3} [and those which follow, such as Eq.~(\ref{4})] give properly scaled dimensions of the image. Dimensions seen through the lens are larger by a factor of the lens magnification.

The Airy intensity in Eq.~(\ref{3}) drops from 1 to 0 at the first zero of the Bessel function, $x \approx 3.83$. This number defines the Airy radius $r_{A}$. Setting $kr_{A}b/f=3.83$, we obtain
\begin{equation}\label{4}
r_{A}=\frac{.61\lambda}{b/f}.
\end{equation}

Because viewing is subjective, the Airy radius may not be perceived as the boundary of the Airy pattern light intensity (the ``Airy disc"), but it is not far off. For consistency with the non-Airy intensity pattern that appears as $b$ is increased, which also falls off rapidly with distance but does not vanish, we define the light boundary to occur at 5\% of the peak value. Because $I_{A}(3.01\ldots)= 0.05$, this criterion puts the radius of the light boundary at $R= (3.01/3.83)r_{A}\approx 0.8r_{A}$.

 \begin{figure}[t]
\begin{center}
\includegraphics[width=0.5\textwidth]{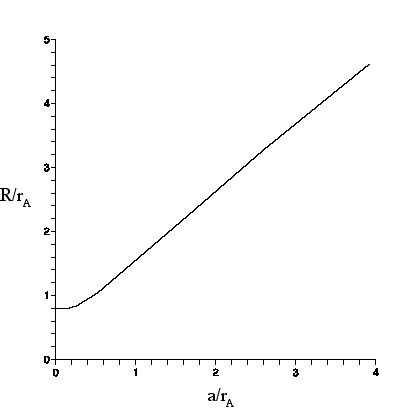}
\caption{\label{F32}For an object circle of radius $a$, R is the image circle's radius, defined as where the intensity is 5\% of the intensity at the center of the image circle. $r_{A}$ is the Airy radius.}  
\label{default}
\end{center}
\end{figure}

As $b$ increases, the Airy radius $r_{A}$ decreases according to Eq.~(\ref{4}), increasing the resolution. Moreover, more light exits the lens, increasing the visibility. As $b$ is increased further, visibility and resolution eventually start to decrease. The light intensity outside $r_{A}$ increases, and the light intensity inside $r_{A}$ decreases due to spherical aberration, that is, rays at the outer edge of the exit pupil come to a focus closer to the lens than do paraxial rays. A design choice called the Strehl criterion\cite{Strehl} suggests an optimal choice of $b$ that keeps spherical aberration at a tolerable minimum while maximizing visibility. This design choice is that the intensity on the optic axis (in the image plane that minimizes the observed disc radius) should be 80\% of $I_{A}(0)$. The intensity shape is still close to the Airy distribution, and hence the image is described as ``diffraction limited," which we shall assume hereafter.

Instead of a point source, consider an extended object, modeled by a circle of radius $a$ illuminated by incoherent light. In geometrical optics, each point on the object plane is imaged onto a point on the image plane for an ideal lens. Therefore, there will be a circular image, which (when account is taken of the lens magnification) appears also to have radius $a$. For an actual lens, each point in the object plane becomes an Airy disc in the image plane. These discs add little spotlights of radius $r_{A}$, with centers uniformly distributed in a circle of radius $a$. Therefore, the image radius $R$ is larger than $a$.

Equation~(\ref{G28}) gives the intensity of the image pattern at any radius in the image plane. Numerical evaluation of Eq.~\eqref{G29} results in Fig.~\ref {F32}, which shows $R/r_{A}$ versus $a/r_{A}$. For $a/r_{A}\lesssim 0.25$, the centers of the Airy discs that contribute to the intensity are so close together that the intensity is close to the Airy pattern. Thus, $R/r_{A}\approx 0.80$ as discussed following Eq.~(\ref{4}). As $a/r_{A}$ increases beyond $\approx 0.25$, the image radius $R$ increases as well, because the Airy disc centers are now spread out over a non-negligible range. From Fig.~\ref {F32} we see that for $a/r_{A}\approx 0.5$ that $R/r_{A}\approx 1$, and for $a/r_{A}\approx 1$, $R/r_{A}\approx 1.5$.

For very large $a/r_{A}$, at the image circle center point or at a point somewhat farther out from the center, the intensity is contributed mostly by Airy discs whose centers lie within an Airy radius of the point. Thus, at the center and to an extent beyond, the intensity remains almost constant but, at $a-r_{A}$ from the center, the intensity starts to drop. At the edge (defined as the circumference of a circle of radius $a$ in the image plane), the intensity is about half that in the center because the edge is nearly a straight line, and there are only Airy discs on the inner side of the edge that contribute. The intensity drops off further as the distance from the center increases beyond $a$, reaching 5\% of $I_{A}(0)$ at $R\approx a +r_{A}$. Thus $(R-a)/r_{A}$ approaches 1 as $a/r_{A}$ becomes very large.

In Fig.~\ref{F32} the largest value shown is $a/r_{A}=4$ at which $(R-a)/r_{A}\approx 0.7$. (If the graph were to be extended, we would find that $a/r_{A}=17, 30$ correspond to $(R-a)/r_{A}\approx 0.9, 0.99$ respectively.) Figure~\ref{F32} will be used to find the actual sizes of spherosomes from their observed sizes.

\subsection{Polystyrene Spheres}

To provide an experimental counterpart to these calculations, slides of $0.3\,\mu$m and $1\,\mu$m diameter polystyrene spheres\cite{Pella} (the standard deviation of the diameters is less than 3\%) were prepared and photographed using the Olympus BX-50 microscope, along with a scale whose line spacing is $\,2\mu$m. The microscope has $r_{A}=0.45\,\mu$m. For $0.3\,\mu$m diameter spheres, because $a= 0.15\mu$m and thus $a/r_{A}= 0.33$ (see Fig.~\ref {F32}), we find that $R/r_{A}\approx 0.86$. Therefore, the spheres should appear with diameter $2R\approx 2(0.86r_{A})\approx 0.77\,\mu$m.

The digital image was enlarged until it appeared as composed of pixels, each a $0.2\,\mu{\rm m}\times 0.2\,\mu$m square. Spheres which stood alone (many spheres cluster) typically appeared as $3\times3$ pixel grids (dark in the middle, and grey on the outside, with the surrounding pixels lighter and more or less randomly shaded), although a $4\times4$ grid for a few could not be ruled out. Thus the diameter of the spheres appeared to be $\approx 0.6\,\mu$m, with an error of a pixel size, consistent with the theory.

For $1\,\mu$m diameter spheres, because $a= 0.5\,\mu$m and thus $a/r_{A}=1.1$, we find that $R/r_{A}\approx 1.7$ from Fig.~\ref {F32}. Therefore, the spheres should appear to have a diameter $2R\approx2(1.7r_{A})\approx 1.5\,\mu$m.

In the unenlarged photograph, isolated spheres seemed to be only slightly larger than $1\,\mu$m, perhaps 1.2--1.3\,$\mu$m, with a bright center (the spheres are transparent) and dark boundary. However, when enlarged so that the pixels can clearly be seen, particularly the outermost light grey ones, the spheres typically appeared as an $8\times8$ grid. Thus the diameter of the spheres appeared to be $1.6\,\mu$m, with an error of a pixel size, consistent with theory.

\subsection{Spherosome Sizes and Brown's Lens}\label{S5D}

We have concluded that the observed spherosome sizes will appear larger than their actual sizes. Moreover, we expect that the spherosome sizes observed by Brown are larger than what we observed with the Olympus microscope due to a larger Airy radius for Brown's lens than the $0.45\,\mu$m Airy radius for the Olympus microscope. The universal size of Brown's ``molecules," regardless of their source, can be attributed to their being small enough so that their Airy disc is what Brown observed.

We obtained an electron microscope picture of amyloplasts (see Fig.~\ref{f100}). We do not have an electron microscope picture of spherosomes to indicate their actual sizes. That is a challenging project for the future. Unlike amyloplasts which are structurally robust, spherosomes are membrane bound lipid droplets. When an attempt is made to concentrate them by filtering so that there are sufficient numbers to view, they coalesce, and appear as an amorphous mass.

We therefore estimated the actual spherosome sizes using the theory and our microscope observations. According to Fig.~\ref{f31}, the diameter of most spherosomes appear to peak at $\approx 1\pm.1\,\mu$m, the largest being perhaps $1.3\pm 0.1\,\mu$m in diameter.

For the smallest spherosomes $R/r_{A}\approx (0.9/2)/0.45\approx1$, for most spherosomes $R/r_{A}\approx (1/2)/0.45\approx1.1$, and for the largest spherosomes $R/r_{A}\approx (1.4/2)/0.45\approx 1.6$. From Fig.~\ref{F32} we see that $a/r_{A}\approx 0.5$, $ 0.6$, and $1.1$ for these three size ranges. Thus, these spherosome radii are $a\approx 0.5\times 0.45\approx 0.2\,\mu$m, $ 0.6\times 0.45\approx 0.27\,\mu$m, and $1\times 0.45\approx 0.5\,\mu$m respectively; that is, the diameters are approximately $0.4\,\mu$m, $ 0.54\,\mu$m and $1\,\mu$m respectively.

Wth these results we try to determine some of the properties of Brown's lens. We assume that the minimum size of his ``molecules" corresponds to the Airy disc, that is, $a/r_{A}<0.3$ in Fig.~\ref{F32} for which $R/r_{A}\approx 0.8$. Because Brown cites the minimum diameter of his ``molecules" as $\approx 1.3\,\mu$m, we have $R\approx 0.65\,\mu$m, and therefore the Airy radius of Brown's lens is deduced to be
\begin{equation}
r_{A}=R/0.8\approx 0.65/0.8 \approx 0.8\,\mu\mbox{m}.
\end{equation}
From Eq.~(4) we conclude that the radius of the exit pupil of his $f=1/32$, $\approx.8$\,mm lens was
\begin{equation}
b=\frac{0.61\lambda f}{r_{A}}=\frac{0.61\times 0.55\times 0.8}{0.8}\approx 0.35\,\mbox{mm}.
\end{equation}

As a consistency check, we note that Brown quoted the maximum diameter of his ``molecules" as $\approx 1.7\,\mu$m. Then, $R/r_{A}\approx(1.7/2)/0.8\approx 1.1$. From Fig.~\ref {F32} we see that this value corresponds to $a/r_{A}\approx 0.6$. We deduce that the actual radius of these largest spherosomes is $a\approx 0.6\times 0.8\approx 0.5\,\mu$m, that is, their diameter is $\approx 1\,\mu$m. This value agrees with the estimate of the actual radius of the largest spherosomes made previously, which was based on observations with the Olympus microscope.

\subsection{Amyloplast Brownian Motion and Rotation}\label{S5E}

\begin{figure}[h]
\begin{center}
\includegraphics[width=0.5\textwidth]{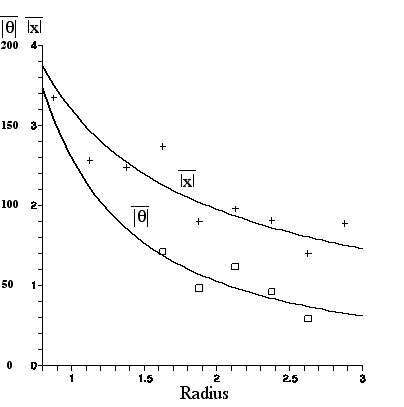}
\caption{\label{f10}Mean linear displacement $\bar{|x|}$ in $\mu$m and mean angular displacement $\bar{|\theta|}$ in degrees, vs R in $\mu$m, for amyloplasts undergoing Brownian 
motion for 60 sec. The least squares 
fit curves depicted here are given in Eq.(\ref{5}). }
\label{default}
\end{center}
\end{figure}

We next analyze the observed Brownian motion of the amyloplasts. In the following, $R$ is half the length of the long axis of an amyloplast.

From Fig.~\ref{f9} the $x$, $y$, and $\theta$-displacements of each amyloplast over a one minute interval were found. Because of the possibility of overall fluid flow (assumed to be constant and irrotational in the region containing the observed particles), the mean displacement was calculated, and found to be 0.053\,$\mu$m in the $x$-direction (negligible flow) and $-0.847\,\mu$m in the $y$-direction. These values were subtracted from each displacement to give the Brownian contribution.

A plot of the mean linear displacement and the mean angular displacement as a function of $R$ is given in Fig.~\ref{f10}. Smallest and largest $R$ values were omitted (which is why there are fewer data points representing $\overline{|\theta|}$ than $\overline{|x|}$) due to the small number of data points. Also shown is the least squares fit to the power law $A/R^{B}$ for each set of data. The results, compared to the predictions given in Eqs.~(\ref{1}) and (\ref{2}) (setting $R_{\rm eff} = R$) are
\begin{align}\label{5}
\overline{|x|} &=\frac{3.2}{R^{0.7}} \quad \mbox{compared to}\overline{|x|}=\frac{4.0}{R^{0.5}},\\
\overline{|\theta|}&=\frac{130}{R^{1.3}} \quad \mbox{compared to}\overline{|\theta|}=\frac{201}{R^{1.5}},\label{5'}
\end{align}
where $R$ is in $\mu$m and $\theta$ is in degrees. The powers in Eq.~(\ref{5}) agree reasonably well, considering that no correction has been made for the ellipsoidal nature of the particles, nor for the fact that the observed amyloplast sizes are larger than the actual sizes. The Brownian motion of ellipsoids, first studied by Perrin, is still of interest.\cite{Perrin}

The numerical coefficients in Eq.~(\ref{5}) differ because the last terms on the right-hand side of Eqs.~(\ref{1}) and (\ref{2}) assume the fluid in which the particles are immersed is water. The amyloplasts move in a fluid that is a mixture of water and the intracellular medium, which emerged with the amyloplasts from the pollen. That is, the measured coefficients are proportional to $1/\sqrt{\eta_{\rm fluid}}$, and the expressions based on Eqs.~(\ref{1}) and (\ref{2}) are proportional to $1/\sqrt{\eta_{\rm water}}$. From Eq.~(\ref{5}) we obtain $\sqrt{\eta_{\rm fluid}/\eta_{\rm water}} = 4.0/3.2\approx 1.3$; from Eq.~(\ref{5'}) this ratio is $201/130\approx 1.5$. These estimates of the fluid viscosity are in reasonable agreement, especially considering the omission of the ellipsoidal correction.

\subsection{Ball Lens Microscope}\label{S5F}

\begin{figure}[t]
\begin{center}
\includegraphics[width=0.5\textwidth]{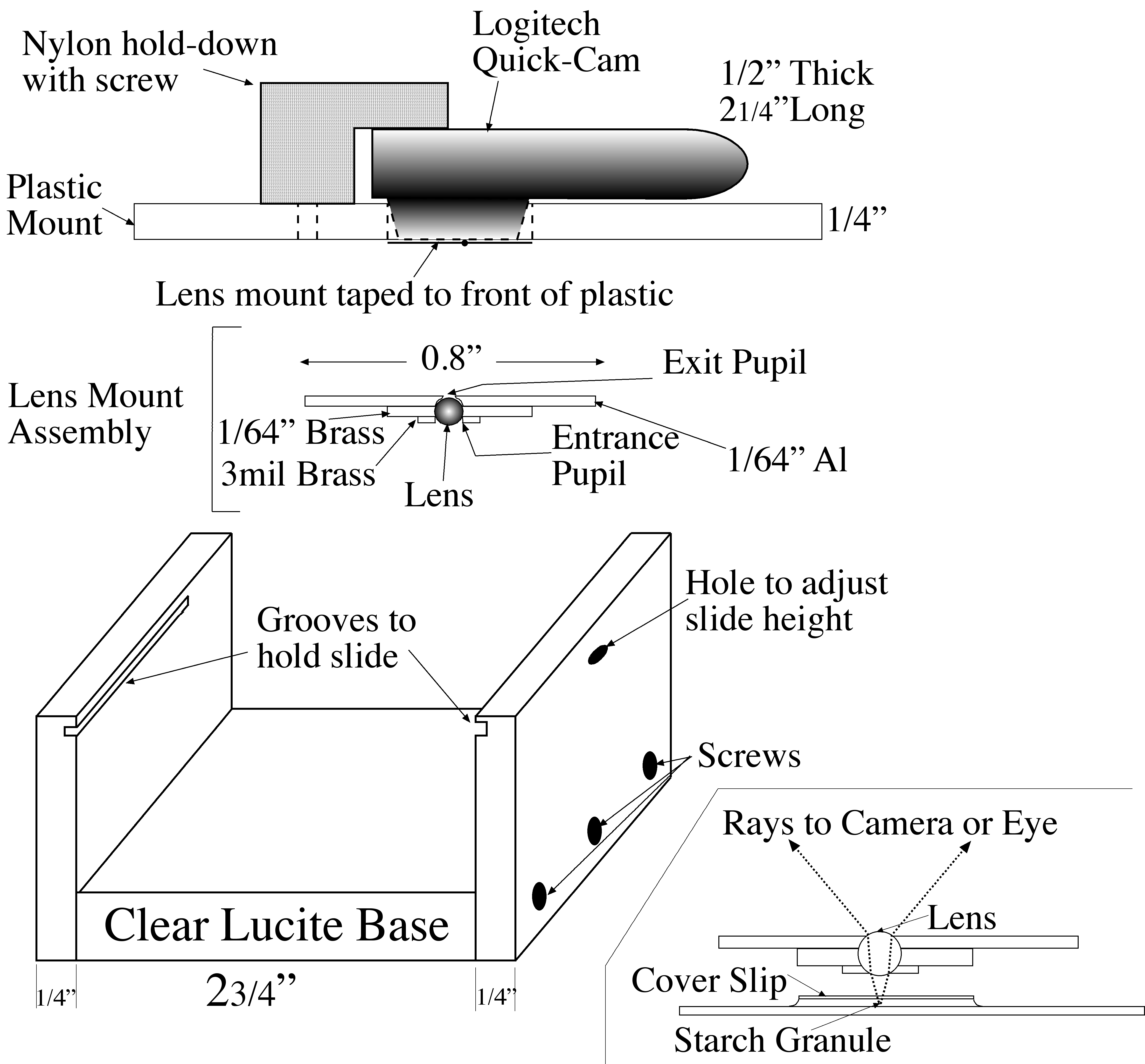}
\caption{\label{f101}Ball lens microscope diagram. The lens mount 
assembly is enlarged.  A ray diagram appears in the inset.}
\label{default}
\end{center}
\end{figure}

Unlike what we did for the spherosomes, we have not presented a theory of how a lens enlarges an object of the shape of the amyloplasts. We discuss here the construction of a single lens microscope with a magnification comparable to Brown's, enabling us to make observations of amyloplast sizes with it and compare them with Brown's observations.

Ground lenses of high magnification are not readily available. However, precision small glass spheres called ball lenses are readily available for use as high magnification lenses.\cite{Ball} We purchased a ball lens of 1\,mm diameter and index of refraction 1.517.\cite{Edmund} The focal length $f$ of a sphere of radius $R$ can be found from the lensmaker's formula\cite{Hardy} for a thick lens of radii $R_{1}$ and $R_{2}$, thickness $T$, and index of refraction $n$:
\begin{equation}\label{6}
\frac{1}{f}=(n-1)\Big[ \frac{1}{R_{1}}- \frac{1}{R_{2}} +\frac{(n-1)T}{nR_{1}R_{2}}\Big].
\end{equation}
Equation~(\ref{6}) with $T=2R$ and $R_{1}=-R_{2}=R$ yields
\begin{equation}\label{7}
f=nR/2(n-1).
\end{equation}
For our lens, $f= 0.733\,{\rm mm}=1/34.6''$, not far from $f=1/32''$ of Brown's lens.

A diagram of the microscope appears in Fig.~\ref{f101}. Details of its construction are given in Appendix~\ref{SD}.

\subsection{Amyloplasts Seen With Ball Lens Microscope}\label{S5G}

\begin{figure}[h]
\begin{center}
\includegraphics[width=0.5\textwidth]{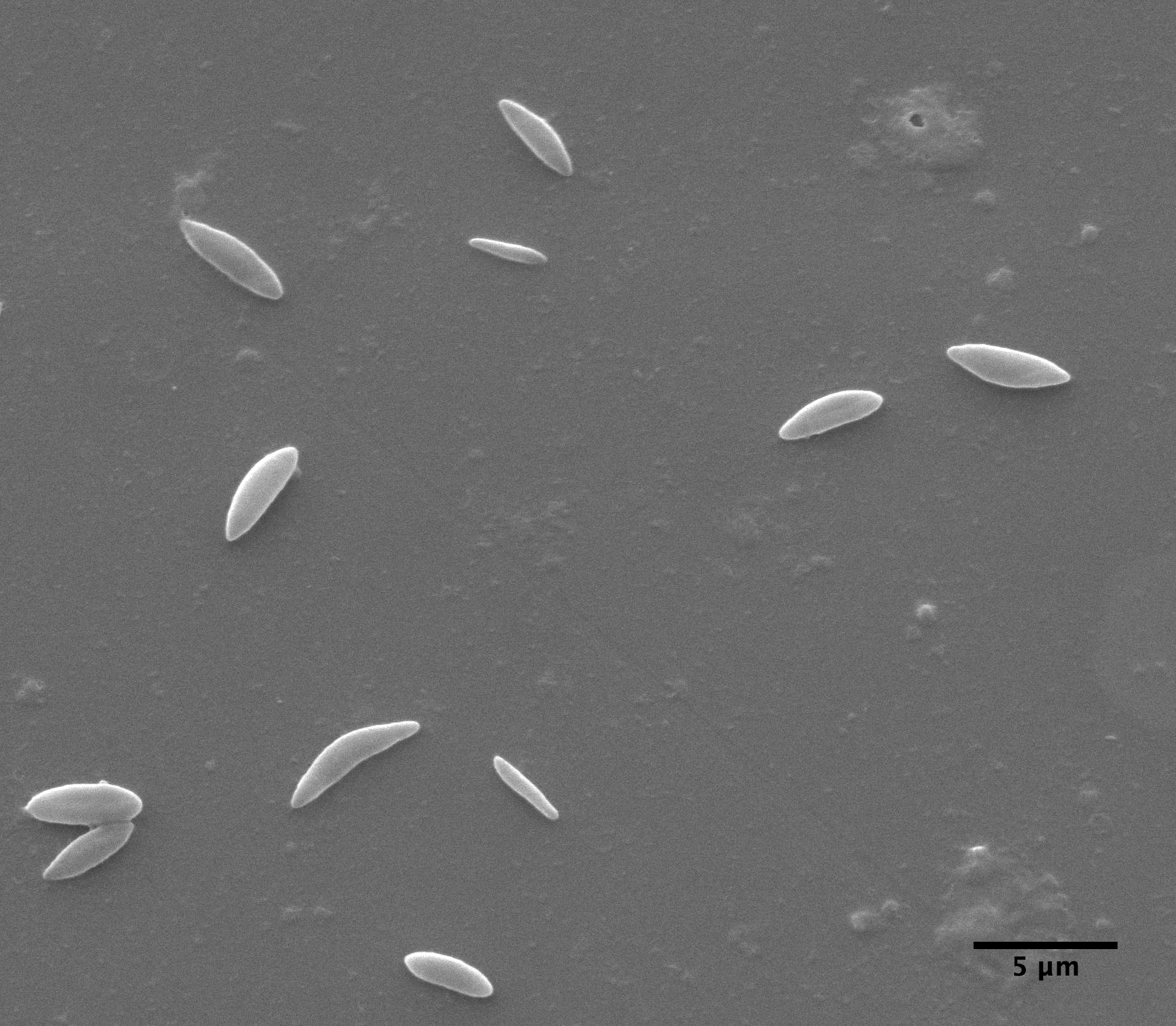}
\caption{\label{f100} Clarkia pulchella amyloplasts photographed with the electron microscope.} \label{default}
\end{center}
\end{figure}

We now address the discrepancy between our observations with the Olympus microscope and Brown's observations with his microscope. Our observations, summarized in Fig.~\ref{f31}, were that the average radius of the amyloplasts are $\approx2\,\mu$m, with the maximum radius $\approx3\,\mu$m. The electron microscope picture of the amyloplasts in Fig.~\ref{f100}, although not depicting a large sample, suggests that the size distribution measured with the Olympus microscope is reasonably accurate. Brown's observations were that their radius range is $\approx3$--4\,$\mu$m. As we will see, our observations with the ball lens are similar to Brown's.

\begin{figure}[h]
\begin{center}
\includegraphics[width=0.5\textwidth]{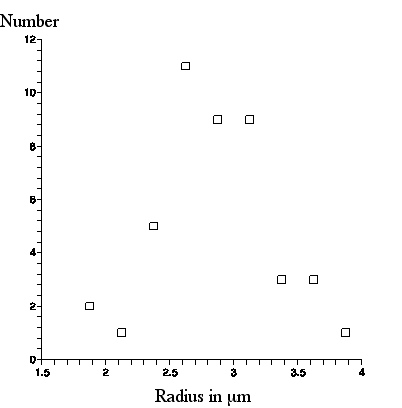}
\caption{\label{f16}Number of amyloplasts in a .25 $\mu$m radius bin vs amyloplast radius (= half amyloplast length) as seen through the ball lens.}
\label{default}
\end{center}
\end{figure}

The Airy radius of the ball lens is $r_{A}=0.61\lambda f/b=0.61 (0.55\,\mu {\rm m})(0.73\,{\rm mm})/0.24{\rm mm}=1.0\,\mu$m. To check that $r_{A}=1\,\mu$m, a slide containing 1$\,\mu$m diameter polystyrene spheres was photographed through the ball lens. Another slide containing a scale with marks 10$\,\mu$m apart was separately photographed and both photographs were superimposed. The image was enlarged so that the pixels could be seen and was analyzed, as described for the spheres photographed with the Olympus microscope. The result was that the diameter of the polystyrene spheres appeared to have diameter $2.1\pm.2\,\mu$m.

For a theoretical comparison, with $a/r_{A}=0.5/1=0.5$, we read from Fig.~\ref{F32} that $R/r_{A}\approx1.1$. Therefore, it is predicted that the apparent radius of the spheres should be $R=1.1r_{A}=1.1\,\mu$m, or diameter 2.2\,$\mu$m, in good agreement with our observation.

	\begin{figure}[h]
\begin{center}
\includegraphics[width=0.5\textwidth]{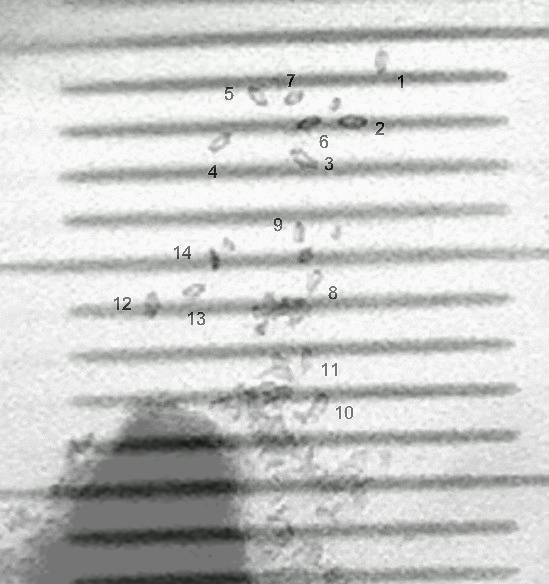}
\caption{\label{f15}Amyloplasts photographed with the ball lens microscope.  The superimposed scale marks are 10$\mu$m apart. }
\label{default}
\end{center}
\end{figure}

We next compare the amyloplast sizes seen with the Olympus microscope and amyloplast sizes seen through the ball lens. Figure~\ref{f15} shows a portion of a photo taken through the ball lens of a slide containing amyloplasts that emerged from a pollen grain (whose out-of-focus edge appears at the lower left).

A photograph of a scale was superimposed and the photograph was further enlarged so that pixels were visible. The radius (half the length) of 44 amyloplasts was measured, 14 of which appear in Fig.~\ref{f15}. A histogram of radii of the amyloplasts is given in Fig.~\ref{f16}. We see that the amyloplasts appear through our ball lens to have an average radius of $\approx 3\,\mu$m, with a maximum radius $4\,\mu$m. This value is about 1\,$\mu$m larger than what was observed with the Olympus microscope (see Fig.~\ref{f31}), but equal to what Brown said about the amyloplast radii he observed through his lens.

This excellent agreement between the observations with our ball lens and Brown's observations with his lens should be tempered by the realization that our lens has $r_{A}\approx1\,\mu$m and exit pupil $b=0.24$\,mm, whereas we have deduced that Brown's lens had $r_{A}\approx 0.8\,\mu$m and exit pupil $b= 0.35$\,mm. However, it leaves little doubt that Brown was seeing enlarged amyloplasts on account of the diffraction and possible spherical aberration of his lens.

\begin{acknowledgments}
It is a pleasure to thank Diane Bilderback, Brent Elliot, Brian Ford, Armando Mendez, Michael Milder, Bill Pfitsch, Hilary Joy Pitoniak, Diana Pilson, Bronwen Quarry, James Reveal, Ann Silversmith, Ernest Small, and David Mabberley for help with this endeavor. The electron microscope photographs were taken by one of the authors, K.B.
\end{acknowledgments}

\appendix

\section{Brownian Translation and Rotation} \label{SA}

A number of places\cite{Morse} exhibit the derivation of Einstein's expression\cite{Einstein} given by Langevin\cite{Langevin} for the mean square distance $\overline{x^{2}}$ traveled in time $t$, by an object of mass $m$ undergoing Brownian motion under a viscous force $-\beta v$ and a random force $f(t)$ satisfying $\overline{f(t)}=0$ which is assumed to be uncorrelated with $x$.  Starting from Newton's second law, 
\begin{subequations}
\label{A1}
\begin{align}
\frac{dx}{dt}&=v\\
m\frac{dv}{dt}&=-\beta v+f(t)
\end{align}
\end{subequations}
\noindent and using the equipartition theorem,
\begin{equation}\label{A2}
\frac{1}{2}m\overline{v^{2}}=\frac{1}{2}kT.
\end{equation}
\noindent one obtains 
\begin{equation}\label{A3}
\overline{x^{2}}=\frac{2kTt}{\beta}.
\end{equation}

For Brownian rotation through angle $\theta$ about an axis, for an object of moment of inertia $I$, Newton's equations are
\begin{subequations}
\label{A4}
\begin{align}
\frac{d\theta}{dt}&=\omega \\
I\frac{d\omega}{dt}&=-\beta ' \omega+\tau(\omega),
\end{align}
\end{subequations}
where the equipartition theorem is
\begin{equation}\label{A5}
\frac{1}{2}I\overline{\omega^{2}}=\frac{1}{2}kT.
\end{equation}
\noindent and $\overline\tau(\omega)=0$. Equations~(\ref{A4}, \ref{A5}) are analogous to Equations~(\ref{A1}, \ref{A2}) so, by the identical argument, one obtains in analogy to Eq.~(\ref{A3}), 
\begin{equation}\label{A6}
\overline{\theta^{2}}=\frac{2kTt}{\beta'}.
\end{equation}

	The values of of $\beta$ and $\beta'$ depend on the shape of the object. For a sphere $\beta=6\pi\eta R$ is given by Stokes law, and it can be shown that $\beta'=8\pi\eta R^{3}$. \cite{website,Huang} The derivation of these expressions is surprisingly complicated. If the shape of the object is not a sphere, we expect that the expression for the force is of the same form as Stokes law with the radius $R$ replaced by an effective radius. Results are available for an ellipsoid, which is the approximate shape of an amyloplast.\cite{website,Huang}

It is useful to have an expression for the mean distance $\overline{|x|}$. It can be argued that the particle position probability density distribution is well approximated by a gaussian distribution,
\begin{equation} \label{A7}
P(x)=\frac{1}{\sqrt{2\pi \overline{x^{2}}}}e^{-x^{2}/2\overline{x^{2}}},
\end{equation}
which yields the result
\begin{equation}\label{A8}
\overline{|x|}=2\int_{0}^{\infty}dxxP(x)= \sqrt{\frac{2}{\pi}{\overline{x^{2}}}}\approx 0.80 \sqrt{\overline{x^{2}}}.
\end{equation}
\noindent Similarly,
\begin{equation}\label{A9}
\overline{|\theta|}= \sqrt{\frac{2}{\pi}{\overline{\theta^{2}}}}\approx 0.80 \sqrt{\overline{\theta^{2}}}.
\end{equation}

\section{Image of a Disc}\label{SC}

We consider the image of a uniformly illuminated hole of radius $a$. The hole models a transparent object such as a spherosome or a polystyrene sphere. We assume that the lens is diffraction limited; that is, the exit pupil has been chosen so that the image of a point source is the Airy intensity distribution, and the hole is illuminated with incoherent light.

In geometrical optics light from each uniformly illuminated point of the object plane passes through the lens and is focused as an illuminated point on the image plane (for an ideal lens). The properly scaled image of all these illuminated points would be a uniformly illuminated circle of radius $a$. We call this circle the ``geometrical image circle," and call its circumference the ``geometrical image circle edge." According to physical optics, diffraction surrounds each imaged point with its own Airy disc, so that the actual image circle extends beyond the geometrical image circle edge. The intensities add, and thus at a point $\bf{r}$ on the image plane, the net intensity is
\begin{equation}\label{G28}
I(r)\sim\frac{1}{\pi}\!\int_{A_{0}}dA_{0}\bigg[\frac{J_{1}(k\tilde{b}|\bf{r}-\bf{r}_{0}|)}{|\bf{r}-\bf{r}_{0}|}\bigg]^{2},
\end{equation}
where $A_{0}$ is the area of the geometrical image circle, and $\tilde{b} = b/f$ is the numerical aperture.

Consider the special case $ r_{A} \ll a$. The intensity at the center point of the image circle is, by Eq.~(\ref{G28}),
\begin{equation}
\label{thiseq}
I(0)\sim\frac{1}{\pi}\!\int_{0}^{a}r_{0}dr_{0}\!\int_{0}^{2\pi}d\phi \bigg[\frac{J_{1}(k\tilde{b}r_{0})}{r_{0}}\bigg]^{2}\approx 1.
\end{equation}
In Eq.~\eqref{thiseq} the limit $a$ has been extended to $\infty$ with no appreciable error, because the major contribution is from Airy discs centered within a distance $r_{A}$ of the origin.

As the point of interest moves off center, the intensity remains essentially constant until a distance $\approx a-r_{A}$ from the center, that is, at a distance $r_{A}$ inside the geometrical image circle edge. Then $I$ starts to decrease, reaching the value of $\approx 0.5$ at the geometrical circle edge.

Now, we turn to the quantitative analysis of the general case with no restriction on the relative sizes of $a$ and $r_{A}$. We shall calculate the intensity outside the geometrical image circle at the center of a coordinate system, $x=y=r=0$ placed at a distance $D$ beyond the geometrical image circle edge. With the image circle in the $x$-$y$ plane, the center of the geometrical image circle in this coordinate system is at $(x,y)=(a+D,0)$. Consider contributing Airy disc centers which lie within the geometrical image circle, between $r_{0}$ ($D\leq r_{0}\leq 2a+D$) and $r_{0} + dr_{0}$, along an arc subtending an angle $2\phi$. The geometrical image circle circumference $(x-a-D)^{2}+y^{2}=a^{2}$ cuts this arc at two points. If we let $x=r_{0}\cos\phi$ and $y=r_{0}\sin\phi$ in this expression, we can find $\cos\phi$ and Eq.~(\ref{G28}) becomes
\begin{equation}\label{G29}
I_{\rm out}(D) \sim \frac{2}{\pi}\int_{D}^{2a+D}r_{0}dr_{0}\cos^{-1}\bigg[\frac{r_{0}^{2}+D^{2}+2aD}{2r_{0}(a+D)} \bigg] \bigg[\frac{J_{1}(k\tilde{b}r_{0})}{r_{0}}\bigg]^{2}
\end{equation}

For completeness, we give the comparable expression for the intensity inside the geometrical image circle. Again, we calculate the intensity at $r=0$, where the origin of this new coordinate system is a distance $D$ away from the center of the geometrical image circle. There are two contributions, one from a circular area of radius $a-D$, and the other from the rest of the disc ($a-D\leq r_{0}\leq a+D$):
\begin{eqnarray}\label{G30}
I_{\rm ins}(D)& \sim&2\!\int_{0}^{a-D}dr_{0}\frac{J_{1}^{2}(k\tilde{b}r_{0})}{r_{0}}\nonumber\\
&&{}
+\frac{2}{\pi}\int_{a-D}^{a+D}dr_{0}\cos^{-1}\bigg[\frac{r_{0}^{2}+D^{2}-a^{2}}{2r_{0}D} \bigg]\frac{J_{1}^{2}(k\tilde{b}r_{0})}{r_{0}} .
\end{eqnarray}
For large $a$, Eq.~(\ref{G29}) becomes
\begin{equation}
I_{\rm out}(D)\approx\frac{2}{\pi}\!\int_{D}^{\infty}dr_{0}\cos^{-1}\bigg[\frac{D}{r_{0}} \bigg]\frac{J_{1}^{2}(k\tilde{b}r_{0})}{r_{0}},
\end{equation}
which is a function of $k\tilde{b}D=3.83(D/r_{A})$. At the edge of the geometrical image circle, $D=0$, and the integral gives $I_{\rm out}(0)\approx 0.5$. Numerical evaluation shows $I_{\rm out}(D)$ drops from 0.5 at $D=0$ to $\approx 0.05$ at $D=r_{A}$. Although it is somewhat subjective, this result suggests that we should take the perceived edge of the image to be located where the intensity is 5\% of its maximum value at the center of the image circle. Thus, due to diffraction, a hole with a large radius $a$ has a larger image circle radius $R\approx a +r_{A}$.

By changing the variable of integration in Eq.~(\ref{G29}) to $r_{0}/a$, we see that the intensity is a function of two variables, $D/a$ and $k\tilde{b}a/3.83=a/r_{A}$. For each value of $a/r_{A}$, we can numerically solve Eq.~(\ref{G29}) for the value of $D/a$ for which $I_{\rm out}(D)=0.05I(0)$. This value of $D$ is $R$, the radius of the image, and $(R/a)(a/r_{A})=R/r_{A}$. A graph of $R/r_{A}$ versus $a/r_{A}$ is given in Fig.~\ref{F32}.

\section{Microscope Construction}\label{SD}

Refer to the diagram of the microscope in Fig.~\ref{f101}. The lens is sandwiched between two perforated supports. One support was made as follows. A circle of $0.8''$ diameter was cut out of a $1/64"\approx 0.4\,$mm thick aluminum sheet. A 0.8\,mm diameter hole was drilled part way through its center, and then a 0.48\,mm diameter hole was drilled all the way through. Thus, the exit pupil radius was constructed to be 0.24\,mm. A small washer was made from a piece of $1/64''$ brass with a 1\,mm diameter hole drilled through it. The holes in the two pieces were aligned, and the pieces secured to each other with Kapton polyimide tape. The ball lens was placed in the resulting hole, supported by the edges of the 0.48\,mm hole and surrounded by the washer.

The second support consisted of a piece of 3mil$\approx$0.076\,mm brass shim stock with a 0.55\,mm diameter hole in the center. It was secured over the lens with Kapton tape to hold the lens in place and serve as the entrance aperture.

The assembled microscope was then mounted with Kapton tape over the entrance aperture of a Logitech QuickCam Pro USB camera. This camera was chosen because the front of its lens lies very close to the surface of the camera, allowing a very small separation between the microscope and the imaging camera. A $1/4'' \times 3''\times5''$ plastic sheet was fashioned, and a hole was drilled through its center, through which the microscope backed by the camera lens protrudes, and the body of the camera rests on the plastic. An inverted-L shaped ``hold-down" bracket was attached to the plastic sheet to hold the camera.

A U-shaped plastic stand of inner horizontal dimension $2.75''$, that is, slightly less than the length of a microscope slide ($3''$), was constructed from $1/4''$ plastic. Horizontal grooves (rabbets) to support the slide were cut in the inner sides of the U just below the top edges. The plastic sheet holding the microscope/camera rests on the top edges of the U, and can be moved freely over the slide.

A small hole was drilled through one side of the U, partly through and partly below the rabbet. Focus adjustment is achieved by placing a small wedge (for example, a toothpick) through the hole and under the slide. As the wedge is moved in and out, it raises and lowers the slide by a fraction of a millimeter. Light from a small microscope illuminator, collimated to a $1''$ beam, is diffusely reflected from a white surface on which the plastic stand sits, through the slide and into the microscope/camera.

The autofocus camera works at the hyperfocal distance. Slightly diverging rays from the ball lens enter the camera lens to be focused a small distance beyond the focal length on the camera's detector surface. If desired, the microscope can be used with one's eye. Replace the plastic camera mount by a $1/64''$ thick aluminum strip of equal length and width, drill a hole in its center of less than $0.8''$ diameter, and tape the lens mount to its underside.


\begin{thebibliography}{99}
\bibitem{vanderpas} Reference~\onlinecite{vanderpas0} claimed that ``\ldots they [Brown's molecules] were approximately of the same size; their diameter varying between 1.26 and 1.6 microns. These statements are not true, BROWN was led to them because he worked with an imperfect lens at the border line of its magnifying power." Although not strictly correct (there was nothing imperfect about Brown's lens, magnification has a value, not a ``borderline"), it seems to be the only statement in the literature which suggests that Brown's lens affected his observations. The purpose of Ref.~\onlinecite{vanderpas0} was to add a name to Brown's list of 10 people who had observed the motion before Brown did. He called attention to observations briefly described in a paper on the usefulness of cover slips in microscopy in 1784 by Jan Ingenhousz.
\end{thebibliography}
\end{document}